\documentclass[aps,pra,twocolumn,superscriptaddress]{revtex4}
\usepackage{amsmath,epsfig,mathrsfs,graphicx,amssymb}

\def\be#1\ee{\begin{equation}#1\end{equation}}

\newcommand{\ba}{\begin{eqnarray} }
\newcommand{\ea}{\end{eqnarray} }

\begin{document}
\title{Fourth moments reveal the negativity of the Wigner function}
\author{Adam Bednorz}
\email{Adam.Bednorz@fuw.edu.pl}
\affiliation{Faculty of Physics, University of Warsaw, Ho\.za 69, PL-00681 Warsaw, Poland}
\author{Wolfgang Belzig}
\affiliation{Fachbereich Physik, Universit{\"a}t Konstanz, D-78457 Konstanz, Germany}
\date{\today}

\begin{abstract}
  The presence of unique quantum correlations is the core of quantum
  information processing and general quantum theory. We address the
  fundamental question of how quantum correlations of a generic quantum
  system can be probed using correlation functions defined for
  quasiprobability distributions. In particular we discuss the
  possibility of probing the negativity of a quasiprobability by
  comparing moments of the Wigner function. We show that one must take
  at least the fourth moments to find the negativity in general and
  the eighth moments for states with a rotationally invariant Wigner function.
\end{abstract}

\maketitle

\section{Introduction}

The von Neumann projective detection scheme \cite{neumann} allows to
measure simultaneously only commuting quantum observables. It is
impossible to define a useful probability of measurement outcomes for
noncommuting observables, like position and momentum, consistent with
a projective measurement.  To circumvent this problem, it is possible to
define a \emph{quasiprobability} -- the Wigner function \cite{wigner},
consistent with the projective measurement but taking sometimes negative
values. Because of the Heisenberg uncertainty principle, the Wigner
function cannot be directly measurable in the von Neumann
sense. However, one can find it by indirect measurements like the homodyne
tomography \cite{hode}, assuming a special detection scheme and using
the inverse
Radon or Abel transformations.  In this scheme it is essential to use
efficient photon counters.  The negativity of the Wigner function, with
the position and momentum replaced by the electric and magnetic field, has
been demonstrated using single photons and other states \cite{lvov,wex}.

Recently, it has become possible to measure the moments of the
microwave electromagnetic field \cite{menzel,wallraff}.  The microwave
frequencies make it difficult to build efficient photon
counters although there exists a proposal \cite{solano}.  Using
instead linear amplifiers leads to a large detection noise and, contrary
to the homodyne detection, one cannot reconstruct the complete Wigner
function.  Hence, it is relevant to ask if we can show the negativity of
the Wigner function having access only to correlations of position and
momentum.

The detection schemes based on linear amplifiers are easier to
interpret in terms of the quantum \emph{weak measurement} \cite{aav}
than in the traditional projective scheme. The interpretation of weak
measurements leads to paradoxes similar to the negative Wigner
function.  In the weak measurement picture, the detector just adds a large
white noise of the outcome. The detection noise contains no
information about the measured system and can be subtracted by
a deconvolution of the resulting probability distribution.  However, the
remaining distribution is a quasiprobability, taking sometimes
negative values. In the case of momentum and position the
quasiprobability corresponds to the Wigner function
\cite{tsang,schleich,quasi}. The deconvolution is in principle
experimentally feasible but since usually one has access only to
correlation functions, it is useful to develop schemes for them to
detect quantum correlations. Generally we have shown previously that
the knowledge of a finite number of moments of a quasiprobability can show its
negativity \cite{quasi}.

In this paper, we discuss how the negativity of the Wigner function
can appear as violation of an inequality involving only finite
moments.  We emphasize that there exist examples violating the
Cauchy-Bunyakovsky-Schwarz inequality by fourth or higher
moments for position and momentum (or second moments for their squares
or intensities) \cite{reid,vog} but only using the Glauber-Sudarshan
P-function \cite{glauber}, whose negativity is much easier to demonstrate,
compared to Wigner function.  We will show in the following that it
is possible to prove the negativity of the Wigner function using only
the fourth moments for position and momentum although the violating state
is complicated.  For rotationally invariant states, mixtures of Fock
states for a harmonic Hamiltonian with
a rotationally invariant Wigner function, we show that one needs at
least the $8^{th}$ moment to show the negativity of quasiprobability.
Additionally, we provide a connection between continuous time-resolved measurement
of position and momentum which permits to find moments of a single-time Wigner function by
deconvolution in a harmonic system.

The paper is organized as follows. We start from recalling basic properties of Wigner function,
including its negativity. Next, we demonstrate violation of classical inequalities by
fourth moment for a special state and eighth moments for a rotationally invariant state.
Finally, we present a way to find moments in continuous measurement by deconvolution
and close with conclusions.

\section{Wigner function}

We begin with the standard construction of the Wigner function.  Let us
define the dimensionless Hermitian position and momentum operators,
$\hat{x}$, $\hat{p}$ with the property $\hat{x}\hat{p}-\hat{p}\hat{x}=[\hat{x},\hat{p}]=i$ ($\hbar$
is incorporated in the definition of $x$ and $p$).  One can always make
a rescaling $\hat{x}\to\lambda_x\hat{x}$,
$\hat{p}\to\lambda_p\hat{p}$ later to recover the proper dimension.  Instead of
$x$ and $p$, in optical systems one can use the electric and magnetic
fields, $E$ and $B$ for a given mode.  The Wigner function is defined as
\cite{wigner},
\begin{equation}
W(x,p)=\int \frac{d\chi d\xi}{(2\pi)^2} e^{-i(\chi x+\xi p)}\mathrm{Tr}e^{i(\chi\hat{x}+\xi\hat{p})}
\hat{\rho}.
\end{equation}
Here $\hat \rho$ is the density matrix of the system.  One essential
property of the Wigner function is that its marginals correspond to
the standard projective probability,
\begin{eqnarray}
&&\rho(x)=\mathrm{Tr}\hat{\rho}\delta(x-\hat{x})=\int W(x,p)dp,\nonumber\\
&&\rho(p)=\mathrm{Tr}\hat{\rho}\delta(p-\hat{p})=\int W(x,p)dx.
\end{eqnarray}
Similar relations can be found for arbitrary linear combinations of
$x$ and $p$, used e.g. to experimentally extract the Wigner function \cite{lvov}.
We stress that, contrary to the $P-$function, the Wigner function definition does not involve
any characteristic harmonic frequency $\omega$. Nevertheless it is convenient to
represent $\hat{\rho}$ in terms of coherent or Fock states defined for a
harmonic oscillator.

The Wigner function can never be measured directly, which follows from
the Heisenberg uncertainty principle and the consistency of quantum
measurements. It cannot be measurable as a probability because it
takes negative values.  To show this we introduce recursively the Fock basis
states $|n+1\rangle=\hat{a}^\dag|n\rangle/\sqrt{n+1}$ for $n=0,1,2,...$ and
$\langle 0|0\rangle=1$, $\hat{a}|0\rangle=0$. They form the complete set of eigenstates of
the Hamiltonian 
\be
\hat{H}=\hbar\omega\hat{a}^\dag\hat{a},\:\hat{a}=(\hat{x}+i\hat{p})/\sqrt{2},\label{osc}
\ee
with the property $[\hat{a},\hat{a}^\dag]=\hat{1}$. In this notation
we have set $m=\omega^{-1}$.
 We have
$\hat{H}|n\rangle=\hbar\omega n|n\rangle$ and \be
W(x,p)=e^{-(x^2+p^2)}(2(x^2+p^2)-1)/\pi \ee with $W(0,0)=-1/\pi<0$ for
$\hat{\rho}=|1\rangle\langle 1|$.  In general, for pure states
$\hat{\rho}=|\psi\rangle\langle\psi|$ the Wigner function is always
somewhere negative unless the state has a Gaussian wavefunction
\cite{hud}.  The directly measurable quantity is $W(x(t))$ which
allows to retrieve the full Wigner function by a numerically difficult
inverse Radon transform (or Abel transform if the state can be assumed
to be rotationally invariant) \cite{hode}.  The statistics needs a lot
of repetitions of the experiment. Moreover, it needs a strong,
projective measurement, excluding any external noise. This condition
is possible to fulfill theoretically but the experimental
implementation relies much on a hardly verifiable detection scheme
(based on efficient photon-counters).  More common is the situation
with some, in many cases even large detection noise, which on the other hand can be
well estimated and subtracted \cite{menzel} as we show in Sec. V. This
latter method makes it impossible, however, to find the whole Wigner
function. In this scheme one can only find a finite number of moments
of the Wigner function. We stress that the moments of the Wigner function are
naturally measured in a continuous weak measurement (and not the P- and
the Q-function). Furthermore, as we show in Sec. V by a suitable
processing of the continuous signal of position \emph{only}, all moments
involving momentum can be extracted as well.

Motivated by the continuous measuring scheme it is natural to pose the following
questions. Can we say something about the negativity of the Wigner
function from the knowledge of a finite number of its moments? What is
the smallest order of the moment to show that the Wigner function is
negative? How complicated must the violating state be? Below we answer
all these questions.

\section{Nonclassical moments -- general case}

An arbitrary probability distribution must satisfy the generic inequality
\be
\langle f^2(x,p)\rangle\geq 0.
\ee
Once the above inequality is violated, the candidate for a probability
must be somewhere negative and fails the test to be a real, positive
probability. 
In the case of the Wigner function we consider the following expansion
\begin{eqnarray}
&&f(x,p)=\sum_{nm}c_{nm}x^np^m=c_{00}+c_{10}x+c_{01}p+\nonumber\\
&&c_{20}x^2+c_{11}xp+c_{02}p^2+c_{30}x^3+\dots
\end{eqnarray}
with real $c_{nm}$ for $n,m\geq 0$.
We say that $f$ is of the order $k$ if there exits a nonzero $c_{nm}$ for $n+m=k$ and 
all $c_{nm}=0$ for $n+m>k$.

We focus on the following task: find the state $\hat{\rho}$ that violates 
\be
\langle f^2(x,p)\rangle_W\geq 0\label{wiv}
\ee
for the $f$ of the lowest possible order.
One can easily show that $k>1$. If $f(x,p)=c_0+c_{10}x+c_{01}p$ then
\be
\langle f^2(x,p)\rangle_W=\mathrm{Tr}f^2(\hat{x},\hat{p})\hat{\rho}\geq 0.
\ee
Next, the case $k=2$ can be reduced by linear transformations on the
phase space, namely  rotating
$(x,p)\to(x\cos\phi+p\sin\phi,p\cos\phi-x\sin\phi)$, squeezing
$(x,p)\to(g x,g^{-1}p)$ and shifting
$(x,p)\to(x+x_0,p+p_0)$, which is justified  in both the classical and
the quantum case, to
$f_a(x,p)=x^2+p^2+c_0$ or $f_b(x,p)=2xp+c_0$. Other cases either reduce to $f_{a,b}$
or will never violate the inequality (\ref{wiv}). It is enough to consider pure states $\hat{\rho}
=|\psi\rangle\langle\psi|$ because violating a mixed state must contain a violating pure state in its
decomposition. Remember that choosing the Fock space of the harmonic
oscillator we do not lose generality, since an arbitrary state
can be written in terms of Fock states even if the Hamiltonian is different from (\ref{osc}).

For $f_a$, one can show that $x^2+p^2=2a^\ast a$ and
\begin{eqnarray}
&&\langle 2a^\ast a\rangle_W=\mathrm{Tr}(2\hat{n}+1)\hat{\rho},\nonumber\\
&&\langle(2a^\ast a)^2\rangle_W=\mathrm{Tr}(4\hat{n}^2+4\hat{n}+2)\hat{\rho},
\end{eqnarray}
with $\hat{n}=\hat{a}^\dag\hat{a}$. Hence, we can consider only the $|n\rangle$ states
but $4n^2+4n+2>(2n+1)^2$ so $f_a$ will never violate (\ref{wiv}).

We have $2xp=i({a^\ast}^2-a^2)$ and
\begin{eqnarray}
\langle ({a^\ast}^2-a^2)\rangle_W & = & \mathrm{Tr}(\hat{a}^{\dag 2}-\hat{a}^2)\hat{\rho},\\
\langle ({a^\ast}^2-a^2)^2\rangle_W & =
&\mathrm{Tr}\left[(\hat{a}^4+\hat{a}^{\dag 4})\hat{\rho}\right]
 -2\langle (a^\ast a)^2\rangle_W.\nonumber
\end{eqnarray}
This gives also the bound $\langle f_b^2\rangle_W\geq -1$ following from
\be
\langle f_b^2\rangle_W+1=\mathrm{Tr}\hat{\rho}(i(\hat{a}^{\dag 2}-\hat{a}^2)+c_0)^2\geq 0.\label{bou}
\ee
We can consider separately $|\psi\rangle$ made of odd states $|2k+1\rangle$
and of even states $|2k\rangle$ because $f_b$ preserves parity, since operators appear pairwise.

The search for the violation of (\ref{wiv}) reduces to the diagonalization of a quadratic form,
parametrized by $c_0$, which is easily accomplished numerically.
The outcome of the search is that we need at least $5$ states of Fock space and the largest violation
of (\ref{wiv}) occurs for $c_0=0$.
The simplest state violating (\ref{wiv}) with $f_b$ has the structure
\be
|\psi\rangle=v_0|0\rangle+v_4|4\rangle+v_8|8\rangle+v_{12}|12\rangle+v_{16}|16\rangle.
\ee
Since $c_0=0$ and $xp$ shifts the state number by $2$, we omit states $|4k+2\rangle$. The coefficients
$(v_0,v_4,v_8,v_{12},v_{16})$ form the eigenvector of
the matrix
\be
\left(\begin{array}{ccccc}
A_0&B_1&0&0&0\\
B_1&A_1&B_2&0&0\\
0&B_2&A_2&B_3&0\\
0&0&B_3&A_3&B_4\\
0&0&0&B_4&A_4
\end{array}\right),
\ee
with the smallest eigenvalue for
\be
\begin{array}{l}
A_k=32k^2+8k+1,\\
B_k=-\sqrt{4k(4k-1)(4k-2)(4k-3)}.
\end{array}
\ee 
The fact that the matrix has a negative eigenvalue follows from
the negativity of its determinant, equal $-10447775$.  The numerical
diagonalization gives $\lambda_{min}=\langle (2xp)^2\rangle_W=-0.036$
and its normalized eigenvector is given by
$(v_0,v_4,v_8,v_{12},v_{16})=(0.973, 0.206, 0.0897,0.042,
0.0161)$. Going to more complicated states allows for an even stronger
violation. However, we stress that our numerical search has shown,
that at least a superposition of 5 Fock states is necessary. Hence, we
draw the important conclusion, that showing the non-classicality of
the Wigner function using only moments requires at least fourth
moments and states involving coherent superposition of at least 5
states.

The violation for a general state is bounded by Eq.~(\ref{bou}).
It is interesting to note, that one can find a state with $\langle f_b^2\rangle_W$ arbitrarily close to $-1$.
We can conveniently describe such a state by its wavefunction in position space
$|\psi\rangle=\int dx\,\psi(x)|x\rangle$, with $\langle x|y\rangle=\delta(x-y)$, $\hat{x}=x$
and $\hat{p}=-id/dx$. We write (\ref{bou}) in the form
\be
\langle f_b^2\rangle_W+1=\int |\phi(x)|^2dx
\ee
and
\be
\phi(x)=(\hat{x}\hat{p}+\hat{p}\hat{x}+c_0)\psi(x)=
(c_0-i-2ixd/dx)\psi(x).
\ee
To get the minimum of $\langle f_b^2\rangle_W$ we should make $\phi(x)$ as small as possible.
However, trying to get $\langle f_b^2\rangle_W=-1$ implies $\phi=0$ and $\psi(x)\sim |x|^{-(1+ic_0)/2}$, which cannot be normalized.
Instead, we take a regularized wave function
\be
\psi(x)=e^{-|x|/2}|x|^{\epsilon-1/2}/\sqrt{2\Gamma(2\epsilon)}\label{vio}
\ee
with $\epsilon>0$ and the Gamma function $\Gamma(z)$. The Wigner function reads
\begin{equation}
W(x,p)=\frac{|x/2|^{2\epsilon}}{\pi\Gamma(\epsilon)}
\left[\frac{\pi J_\epsilon(2xp)}{|xp|^\epsilon e^{|x|}}+\mathrm{Re}\frac{2K_\epsilon(|x|+2ixp)}{(ipx+|x|/2)^\epsilon}
\right]
\end{equation}
with $W\sim J_0(2xp)-Y_0(2xp)$ for $\epsilon\sim 0$ and $J$,$K$,$Y$ denoting Bessel functions.
We get $\phi(x)=-i(2\epsilon-|x|)\psi(x)$, which gives
$\langle (2xp)^2\rangle_W=2\epsilon-1$. This result shows that we can get arbitrarily
close to the limiting value $-1$ with $\epsilon\to 0$, however, the case
$\epsilon=0$ is impossible as the wave function would not be
normalizable. We stress that (\ref{vio}) represents only an example of a state
with $\langle f_b^2\rangle_W\to-1$ and one can certainly find also such states in the family of
analytic functions, e.g. a sequence of states in finite Fock spaces.

\section{Nonclassical moments at rotational invariance}

We ask the further question: What is the lowest order of $f$ that violates (\ref{wiv})
for rotationally invariant states (considering the Wigner function in
phase space)? The rotational invariance is intimately connected with
the harmonic dynamics 
(\ref{osc}), which induces a rotation frequency $\omega$. It is important since it is often difficult
to control experimentally the phase of the photonic state,
corresponding to the polar angle in the phase space. If the phase is
smeared out then 
the state is rotationally invariant, described by a mixture of Fock states $|n\rangle$, namely
$\hat{\rho}=\sum_n p_n|n\rangle\langle n|$. Again,
it is enough to focus on pure Fock states. Next, the parts of $f$ with different parity
with respect to $x$ or $p$ can be considered separately as their cross products
integrate out to zero for Fock states. The function $f_b$ reduces to $f_a$ for Fock states
so it cannot violate (\ref{wiv}). The only nontrivial case with $k=3$ is then
\be
f_c(x,p)=x(x^2+p^2)+(c_{30}-1)x^3+c_{10}x.
\ee
One can find that for a rotationally invariant distribution
\begin{eqnarray}
&&\langle f^2_c\rangle=(5c_{30}^2+2c_{30}+1)\langle r^6\rangle/16\nonumber\\
&&+(3c_{30}+1)c_{10}\langle r^4\rangle/4
+c_{10}^2\langle r^2\rangle/2,
\end{eqnarray}
where $r^2=x^2+p^2$. The straightforward analysis of the above quadratic form
gives its minimum located at
\begin{eqnarray}
&&c_{30}=\frac{3\langle r^4\rangle^2-2\langle r^6\rangle\langle r^2\rangle}
{10\langle r^6\rangle\langle r^2\rangle-9\langle r^4\rangle^2},\;
c_{10}=\frac{-\langle r^4\rangle\langle r^6\rangle}
{10\langle r^6\rangle\langle r^2\rangle-9\langle r^4\rangle^2},\nonumber\\
&&\langle f_c^2\rangle=
\frac{\langle r^6\rangle(\langle r^6\rangle\langle r^2\rangle-\langle r^4\rangle^2)}
{2(10\langle r^6\rangle\langle r^2\rangle-9\langle r^4\rangle^2)}.
\end{eqnarray}
On the hand, for the Fock state $|n\rangle$ one finds $\langle r^2\rangle_W=2n+1$,
$\langle r^4\rangle_W=4n^2+4n+2$ and $\langle r^6\rangle_W=8n^3+12n^2+16n+6$
so $\langle r^6\rangle_W\langle r^2\rangle_W-\langle r^4\rangle_W^2=12n^2+12n+2>0$.
Hence, $\langle f_c^2\rangle$ will be never negative for Fock states.

The negativity appears for $k=4$, namely for $f_d(x,p)=r^4+c_{20}r^2+c_0$, taking
the state $|1\rangle$.
One gets
\be
\langle f_d^2\rangle_W=10c_{20}^2+c_0^2+6c_{20}c_0+84c_{20}+20c_0+216.
\ee
The minimum of the above function is located at $c_{20}=-12$, $c_0=26$ and equals $\langle f^2_d\rangle_W=-28$.
This is our second result: we need the eighth moments and not less to show
the nonclassicality of rotationally invariant states.

\section{Moments from continuous measurement}

The concept of the Wigner function is
very general as it only relies of the fact that two observable do not
commute. However, in many real experiment the observables are measured
continuously and, hence, it necessary to extend the concept of a
quasiprobability to continuous time-resolved measurements. To this
end, we define a \textit{time-extended} Wigner functional, in accord with
general considerations \cite{tsang,schleich,quasi}, by a path integral
\begin{eqnarray}
\mathcal W[x,p] & = & \int D\chi D\xi\;
e^{S[\chi,\xi]-\int idt[\chi(t)x(t)+\xi(t)p(t)]},\nonumber\\
e^{S[\chi,\xi]} & = & \mathrm{Tr}\mathcal T
e^{\int idt[\chi(t)\hat{x}(t)+\xi(t)\hat{p}(t)]/2}\;
\hat{\rho}\;\times\label{wiext}\\
&&\tilde{\mathcal T}
e^{\int idt[\chi(t)\hat{x}(t)+\xi(t)\hat{p}(t)]/2},\nonumber
\end{eqnarray}
where $\mathcal T(\tilde{\mathcal T})$ denotes (anti)time ordering,
$\hat{A}(t)$ denotes an operator in the Heisenberg picture and $\hat{\rho}$
is the initial density matrix. Such a functional can be measured as
 a convolution with some detection noise \cite{quasi} \be
 \rho[x,p]=\int Dx'Dp'
\rho_d[x-x',p-p']\mathcal W[x',p'],\label{conv} \ee where $\rho_d$
corresponds to the large detection white noise, independent of the
state of the system.  This corresponds to the experimental detection
scheme, in which moments of $\mathcal W$ can be found by deconvolution
\cite{menzel,vog1}.  The measurable quantities are the cumulants \be
\langle\langle A_1\cdots
A_n\rangle\rangle=(-i)^n\left.\frac{\partial^n
    S(\chi_1,\dots,\chi_n)}{\partial\chi_1\cdots\partial\chi_n}\right|_{\chi=0},
\ee and the moments \be \langle A_1\cdots
A_n\rangle=(-i)^n\left.\frac{\partial^n
    e^{S(\chi_1,\dots,\chi_n)}}{\partial\chi_1\cdots\partial\chi_n}\right|_{\chi=0},
\ee where the cumulant generating function (CGF) is defined \be
S(\chi_1,\dots,\chi_n)=\ln\langle
e^{i(A_1\chi_1+\dots+A_n\chi_n)}\rangle.  \ee Knowing the cumulants of
$\rho$ and $\rho_d$ one can find the cumulants and moments of $\mathcal W$
due to the relation between the CGFs \be S_\rho=S_d+S_{\mathcal W}.  
\ee

Now, let us focus on a harmonic oscillator described by the Hamiltonian (\ref{osc}). 
In the Heisenberg picture $\hat{a}(t)=e^{-i\omega t}\hat{a}$.
Due to the fact that all operators in (\ref{wiext}) commute to $c$-numbers, one can now write
\be
e^{S[\chi,\xi]}=\mathrm{Tr}
e^{\int idt[\chi(t)\hat{x}(t)+\xi(t)\hat{p}(t)]}\;
\hat{\rho}.
\ee
One can see also that the Wigner functional (\ref{wiext}) vanishes outside classical trajectories, namely
\be
(da(t)/dt+i\omega a(t))W[x,p]=0,\:a=(x+ip)/\sqrt{2}.\label{evol}
\ee
It leads to a simple picture: once the initial distribution $W(x,p)$ is known, the
general functional $\mathcal W[x,p]$ is obtained by evolving along classical trajectories.
Therefore, for a harmonic Hamiltonian, the complete Wigner functional is
not necessary for the theoretical calculations 
as it reduces to the single-time Wigner function. However, the
classical evolution is helpful for the experimental detection schemes.
It is worth to collect all data from a continuous (weak) monitoring of $x(t)$ and/or $p(t)$ and construct 
the initial Wigner function by a reverse mapping of the classical dynamics.

Let us introduce the notation for the marginal distributions
\begin{eqnarray}
&&\mathcal W[x]=\int Dp\mathcal W[x,p],\:\:\nonumber\\
&&
\mathcal W(x_1,\dots,x_n,p_1,\dots,p_m)=\int DxDp\;\mathcal W[x,p]\times\nonumber\\
&&\delta(x_1-x(t_1))\cdots
\delta(x_n-x(t_n))\times\\
&&\delta(p_1-p(t'_1))\cdots\delta(p_m-p(t'_m)).\nonumber
\end{eqnarray}
We shall switch the notation $x_k\to x(t_k)$ when it is unambiguous.
It is easy to show that
\be
\mathcal W[x(0),x(\pi/2\omega)]=\mathcal W[x(0),p(0)]=W(x,p),
\ee
so one can obtain the usual Wigner function also from only position
measurement,  but at different times, as $p(0)=x(\pi/2\omega)$ follows
from the equation of motion (\ref{evol}). More generally, $W(x,p)$ can be
obtained also from 
the appropriate Fourier components, namely
\begin{eqnarray}
&&W(x,p)=\int Dx'\;\delta\left(x-\int_0^{t_0} 2\cos(\omega t)x'(t)/t_0\right)\times\nonumber\\
&&\delta\left(p-\int_0^{t_0} 2\sin(\omega t)x'(t)/t_0\right)\mathcal W[x'],
\end{eqnarray}
where $t_0\gg\omega^{-1}$ is the total averaging time. 
We stress that the averaging time is limited by the backaction of the
detector, which protects -- by means of the detection noise --  
the Wigner function from its exact measurement in a single run of
an experiment.  

\section{Conclusion}

We have demonstrated that the negativity of the Wigner
function can be shown using only its moments but at least of fourth
order.  As also shown in other context we confirmed that fourth order
correlations are necessary to show nonclassicality of quantum
mechanics \cite{quasi}.  We believe this is fundamentally important
for probing nonclassicality by means of weak measurement, when one has
access only to moments. We are convinced that our results will stimulate an
effort to measure higher order correlations in optical cavities by
means of linear detectors and amplifiers even if they add large
detection noise.

\section*{Acknowledgment}

We acknowledge financial support from the DFG through SFB 767
\textit{Controlled Nanosystems} and SP 1285 \textit{Semiconductor
  Spintronics}.

\end{document}